# Hidden Information in Force Curves: Transient- and Brownian-Driven Dynamics

Sergio Santos[1] and Roger Proksch[2*]
[1]nomos association, Barcelona, Spain; [2]Asylum Research, Oxford Instruments, Santa Barbara, CA, USA
*Correspondence: roger.proksch@oxinst.com

## Abstract

Force–distance curves (FCs) are direct probes of tip–sample interactions in atomic force microscopy, but their dynamic content is often suppressed by filtering, averaging and quasi-static analysis. Indeed, in the forty years since the invention of the AFM, this hidden dynamic signal has been largely ignored, with exceptions restricted to the large oscillations associated with relatively large adhesive separations (snap from contact) or to time-averaged approaches that carry their own challenges. Using high-bandwidth interferometric detection, we directly measure fast FCs, including short-lived cantilever oscillations during snap-in, contact and pull-off. Two excitation mechanisms coexist: thermal fluctuations and transients generated by rapid changes in the tip–sample boundary condition. Cycle-resolved frequency and amplitude estimators provide time-local observables of the evolving cantilever–sample system within single FCs, creating a pixel-by-pixel mechanical-contrast channel that requires no external broadband excitation, resonant tracking or multi-pass imaging. A state-dependent harmonic-oscillator model incorporating van der Waals attraction, hysteretic capillary-bridge formation and rupture, hydration-mediated damping and repulsive contact stiffness partitions each FC into five interaction regimes and reproduces the observed responses. In repulsive contact, frequency shifts yield stiffness maps consistent with contact-mechanics estimates, enabling high-throughput stiffness mapping and localization of transient tip–sample events from individual pixels. More broadly, this hitherto hidden signal may pave the way for new ways of studying and exploiting Brownian motion in a variety of systems, providing the capability to follow time-resolved (non-)Markovian and (non-)ergodic dynamics mediated by the tip–sample interaction, including single-event dynamics.

## Introduction

Since the invention of the atomic force microscope (AFM),[1] force–distance curves (FCs) have played a central role in probing tip–sample interactions and local mechanical properties at the nanoscale.[2] In a FC, the relative tip–sample separation is ramped while the cantilever deflection is recorded, yielding a direct measurement of the interaction force $F$, through Hooke's law $F = kz$, where $k$ is the cantilever spring constant and $z$ is the displacement. The approach and retract trajectories encode long-range forces (van der Waals, electrostatic, magnetic, and capillary[3] interactions), short-range repulsive contact forces related to elasticity and plasticity,[4] adhesion, and non-conservative dissipation manifested as hysteresis and creep.[2,5,6]

For much of the history of AFM, FCs were acquired slowly (order one curve per second), limiting their utility as an imaging mode. The development of Pulsed Force Mode[7] and related fast force mapping techniques dramatically increased acquisition speed and enabled spatial mapping of stiffness and adhesion-related contrast in reasonable time frames.[8,9,10,11,12,13,14] Increasing force-curve speed, however, brings cantilever dynamics to the forefront. Impulsive events associated with Brownian forcing and other rapid changes in the tip–sample boundary conditions excite oscillations whose decay occurs on microsecond-to-millisecond timescales. In principle, these oscillations contain useful information, yet these hidden dynamics have been largely ignored in the forty years since the AFM was invented. The few exceptions[15,16] have been largely restricted to the large oscillations associated with relatively large adhesive separations (snap from contact) or to time-averaged approaches such as that of Eppell et al.,[9] which carries its own challenges, as discussed below. Otherwise, the common approach is to treat these oscillations as noise and suppress them by filtering. Here, we show that they represent the direct dynamical response of the coupled cantilever–sample system and distinguish thermal Brownian motion from boundary-condition-induced transient motion.[17]

In this work, we acquire high-bandwidth force curves using quadrature phase differential interferometry (QPDI). The low displacement noise floor of $\approx 5\, fm/\sqrt{Hz}$ (i) renders small cantilever motion directly observable throughout the modulation cycle and (ii) provides the necessary precision to sufficiently control the probing force to optimize the resolution with which a given interaction can be explored. To monitor, disentangle and quantify the transient and fast decaying phenomena we develop a "single-cycle" estimator for the oscillator amplitude and frequency that identifies peak to valley time-periods. With this setup we can capture information-bearing measurement-broken ergodicity. Here the observation window is shorter than the characteristic relaxation or transition time encoded in the mechanisms underlying the ergodic mean. Averaging only increases precision relative to the persistent phenomena encoded in that mean. However, our estimator captures information that is otherwise lost to the mean and that only short-window transients can carry. This includes history-dependent transitions with no single-valued mean, fast-decaying relaxation whose transient shape encodes information absent from the converged mean, and singular events that occur only a finite number of times within the observation and for which no ensemble average could ever be formed at all. In short, we treat transient events, Brownian fluctuations and general non-repeating, non-ergodic phenomena as measurable and quantifiable signals, opening the path to high resolution fast-imaging of pixel-by-pixel contrast channels of single event phenomena.

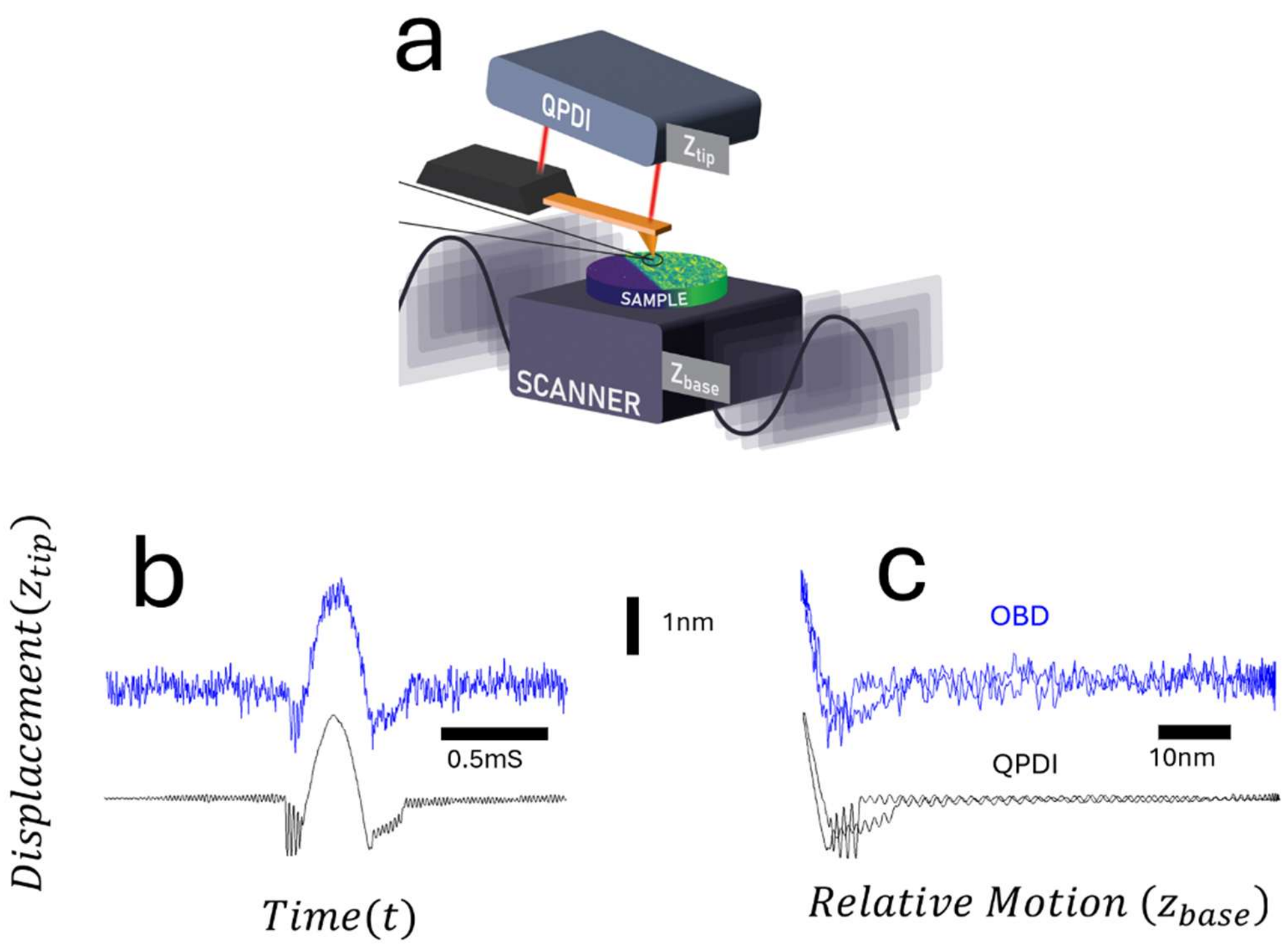

***Figure 1(a)*** *Schematic of the experimental setup showing the scanner base position,* $z_{base}$ *and the cantilever tip position,* $z_{tip}$*, measured during pulsed-force operation. The indentation is calculated from the relative motion between the scanner base and the cantilever tip, indentation* $\delta = z_{base} - z_{tip}$*.* ***(b)*** *Representative displacement time trace from a single pulsed-force modulation cycle comparing the optical beam deflection signal, OBD, with the quadrature phase differential interferometry signal, QPDI. The scale bars correspond to 1 nm in displacement and 0.5 ms in time.* ***(c)*** *Representative displacement plotted as a function of relative scanner motion, indentation* $z_{tip}$*, comparing the OBD and QPDI measurements. The OBD trace is shown in blue and the QPDI trace in black, highlighting the improved displacement resolution of QPDI during the tip–sample interaction. The horizontal scale bar corresponds to 10 nm.*

The experimental setup is illustrated in Figure 1a. Here, $z_{\text{base}}$ is measured by the scanner z-sensor, as the sample is sinusoidally modulated along the z-axis, and $z_{\text{tip}}$ is measured by the QPDI detector. Figures 1b and 1c show representative force profiles illustrating how the data obtained for $z_{\text{base}}$ and $z_{\text{tip}}$ differ according to whether we employ a QPDI detector or the standard OBD signal to monitor single event non repetitive phenomena. These figures show a pulsed force displacement versus time and versus relative motion between the cantilever and sample, respectively. The OBD signal (with a noise floor of $\approx 150\, fm/\sqrt{Hz}$) is shown in blue and the QPDI interferometer signal, (with a noise floor of $\approx 5\, fm/\sqrt{Hz}$) is shown in black. Figures 1b and c display oscillating signals in the QPDI data that are, for the most part, invisible in the simultaneous OBD measurements. Eppell et al.[9] already attempted to resolve the "unexplored" portions of OBD-measured AFM force curves by averaging many curves made over a particular region (as many as 500) to show that there is real nanomechanical information hidden in the noise. In their case, the transition from van der Waals dominated to capillary dominated interaction excited a damped oscillation that was hidden below the noise in single force curves but became visible after registering and averaging multiple curves; the oscillation frequency and damping then provided contrast related to local stiffness and viscosity.

An early and successful example of putting this oscillatory information to use is the "ringing mode" introduced by Dokukin and Sokolov.[15] By analyzing the resonant ringing of the cantilever after the probe detaches from the sample surface—a signal otherwise filtered out or ignored in sub-resonance tapping modes—ringing mode extracts up to five additional compositional channels, including the restored adhesion, disconnection energy loss and disconnection distance. This work established that the post-detachment transient carries genuine nanomechanical information rather than noise. It operates, however, exclusively on the post-detachment portion of the force curve—the rupture-and-recovery we define as regime R5 in the segmentation introduced below—and in the limit of large oscillations excited by relatively large adhesive separations (snap from contact), where the energy transferred to the cantilever amounts to several hundred eV (in contrast to the thermal energy of $\approx 25 meV$) and is readily detectable even with conventional optical beam deflection. The resolution demonstrated in the present work, in comparison, extends down more than two orders of magnitude below the p-p thermal amplitude at room temperature ($A_{p-p} \approx \sqrt{2k_BT/k}$), giving access to both deterministic transients generated by changing tip–sample boundary conditions and phenomena associated with stochastic (Brownian) and non-ergodic events. While the improved temporal requires estimators appropriate for short, non-stationary signals, these measurements capture the dynamics experienced at each pixel, without requiring ensemble averaging. Furthermore, contrast images might be produced where true instantaneous and non-ergodic phenomena can be directly quantified and contrasted with pixel resolution.

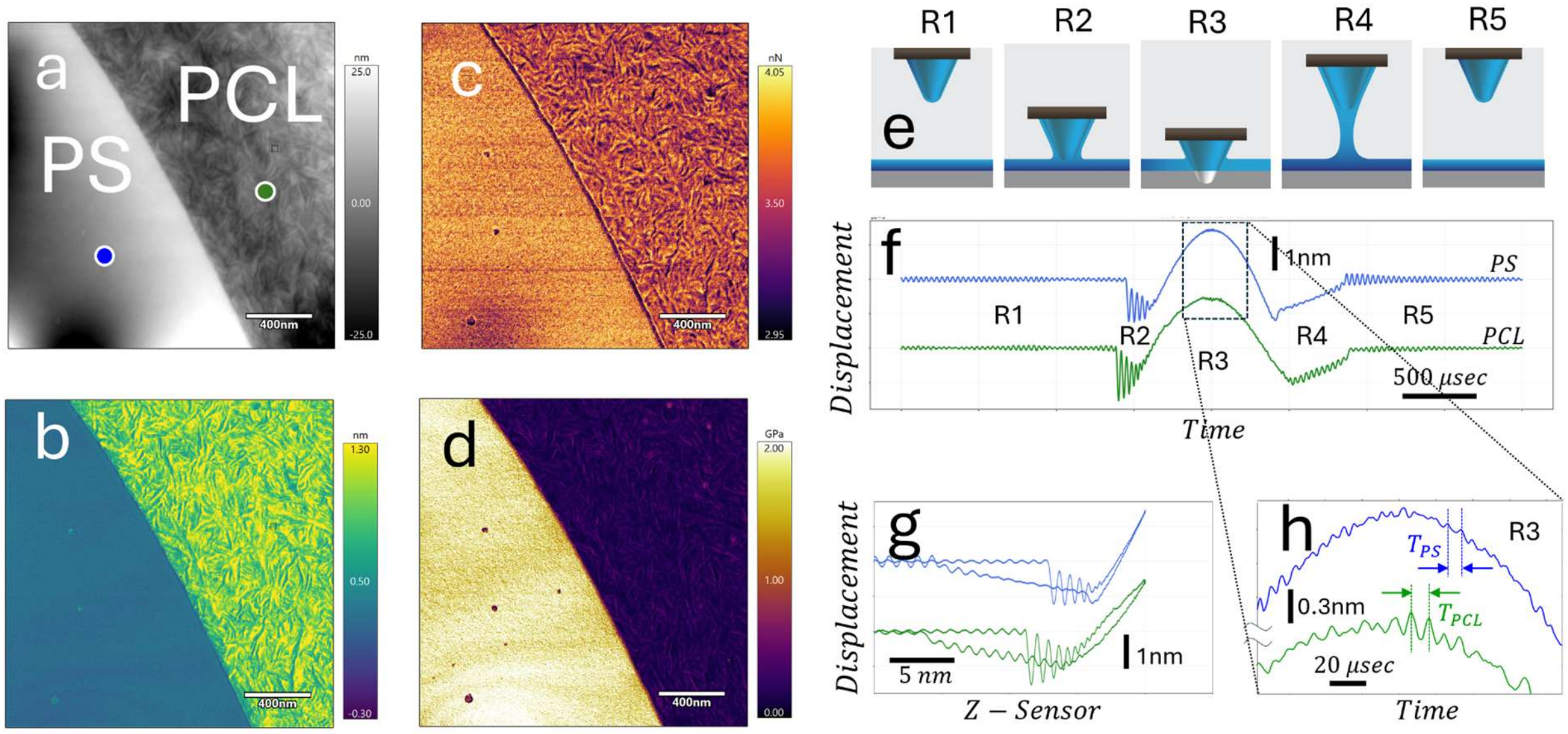

*Figure 2.* (a) Height/topography image of the PS/PCL blend over a 1.5 × 1.5 μm region, showing the mechanically distinct polystyrene (PS) and polycaprolactone (PCL) phases. Representative PS and PCL measurement locations are indicated by blue and green circles, respectively. (b) Spatially resolved maximum indentation map from the same region, highlighting the contrast in tip–sample deformation between the two phases. (c) Corresponding adhesion map showing phase-dependent variations across the PS/PCL interface. (d) Elastic modulus map, showing the strong mechanical contrast between the stiffer PS-rich region and the softer PCL-rich region. Scale bars in (a–d) are 400 nm. (e) Schematic illustration of the five tip–sample interaction regimes, R1–R5, corresponding to noncontact approach, capillary/meniscus formation, maximum indentation, withdrawal, and recovery. (f) Individual displacement versus time traces acquired at the marked PS (blue) and PCL (green) locations, with the five interaction regimes (R1–R5) labeled. Curves are vertically offset for clarity; the vertical scale bar corresponds to 1 nm and the horizontal scale bar to 500 μs. (g) The same events plotted as cantilever displacement versus Z-sensor position, showing the full approach and retraction branches for each material. Vertical and horizontal scale bars correspond to 1 nm and 5 nm, respectively. (h) Magnified view of the maximum-indentation regime (R3) from the dashed region in (f), revealing the material-

dependent oscillation periods, $T_{PS}$ and $T_{PCL}$, of the tip response on each phase. Vertical and horizontal scale bars correspond to 0.3 nN and 20 μs, respectively.

Figure 2 summarizes the spatially resolved mechanical response of the PS/PCL blend and introduces the time-domain segmentation used throughout the remainder of this work. Figures 2a–d show a 1.5 × 1.5 μm region of a spin-coated polystyrene/polycaprolactone film characterized with conventional nanomechanical analysis. Figure 2a shows the sample topography of the polystyrene (PS) and polycaprolactone (PCL) domains, with representative, color-coded measurement locations marked by circles (PS in blue, PCL in green; this color convention is maintained in panels f–h). Figures 2b–d show mechanical-property maps from the same region: maximum indentation (using the definition $\delta(t) = z_{\text{base}}(t) - z_{\text{tip}}(t)$), adhesion, and tip–sample stiffness, respectively. These maps were generated using the conventional analysis tools available in the instrument software and serve as a baseline comparison for the higher-bandwidth, time-resolved analysis developed later in the paper. As expected, the softer PCL region exhibits larger indentation and lower stiffness than the PS region, and the semicrystalline lamellar texture of the PCL phase is clearly resolved in the indentation and adhesion channels. Quantitative modulus determination remains limited by uncertainties in tip radius, contact geometry, adhesion, and the cantilever boundary condition; accordingly, we focus primarily on effective tip–sample stiffness (Figure 2d) rather than absolute modulus.

Figure 2e schematically defines the five tip–sample interaction regimes into which we segment each modulation cycle: (R1) noncontact approach, (R2) capillary/meniscus formation and snap-in, (R3) maximum indentation, (R4) withdrawal with a stretched capillary bridge, and (R5) rupture and recovery to the noncontact state. This segmentation is central to the analysis that follows,

because—as discussed above—the approach and retraction branches constitute memory-dependent states of a non-ergodic mechanism: the capillary bridge present during withdrawal (R4) has no counterpart on approach (R1), so the two branches encode distinct physical information that would be conflated by averaging.

Figure 2f shows the individual interferometric displacement signals versus time for the marked PS and PCL locations from Figure 2a, with the five regimes labeled directly on the traces; the PCL trace is vertically offset for clarity. This time-domain representation makes explicit that the measured signal is not a quasi-static force curve. Instead, the cantilever undergoes transient oscillations whose frequency, amplitude, and decay depend on the instantaneous tip–sample interaction. This is most evident at the R1→R2 transition, where snap-in launches a pronounced ringdown that decays over several oscillation periods, and again in R4, where pull-off excites a second transient as the capillary bridge stretches and ruptures. Notably, the character of these transients differs between the two phases: the contact-coupled oscillations in R2–R4 occur at a higher frequency and decay differently on PS than on PCL, reflecting the different tip–sample stiffness and damping of each material. These short-lived events are precisely the class of information identified above as being suppressed by averaging—their kernel shape resides entirely in the transients.

Figure 2g replots the same events versus Z-sensor position, i.e., in the conventional force-curve representation, showing the full approach and retraction branches for each material. In this domain, the hysteresis between the branches—the signature of the non-ergodic capillary mechanism—is directly visible, as are the superimposed oscillations that a low-pass-filtered, pixel-level analysis would discard. Comparing Figures 2f and 2g underscores a central point of this work: the two representations contain the same raw data, but the time domain resolves the sequence and decay of transient events, while the Z-sensor domain resolves branch-dependent hysteresis; conventional mechanical maps (Figures 2b–d) compress both into a handful of scalars per pixel.

Figure 2h shows a magnified view of the maximum-indentation regime R3, from the dashed region in Figure 2f. At this magnification, a material-dependent time delay between the drive and the tip response is resolved, denoted $T_{PS}$ and $T_{PCL}$ for the two phases, with the condition $T_{PCL} > T_{PS}$ clearly visible in Figure 2h. These time period measurements constitute a direct, single-cycle observable of the local, Brownian-driven tip-sample interaction. Importantly, this quantity is extracted from a single time-resolved cycle rather than from an ensemble or cycle average, illustrating concretely how measurement-broken ergodicity gives access to material-specific information—here, a stiffness—that survives in each individual event but would be smeared or lost in the averaged curves of conventional analysis.

In the following, we will broadly discuss some of the observations of the new information revealed and a general approach to modeling the cantilever dynamics as the tip explores five interaction regimes (R1–5), 1 approach, 2 snap to contact, 3 maximum indentation, 4 withdrawal, and 5 recovery. Then, we will focus in much more detail on two specific domains: the formation and collapse of a meniscus between the tip and the sample (regions R2 and R4) and the mechanical contact region (R3).[18] We note that ambient force interactions are typically discussed as three regions[19] (our regions R1-R3). R4 is not discussed either because the information is missing or because R2 and R4 are conflated. The distinction between R1 and R5 is more subtle since the system is assumed to, because of ergodicity, produce the mean behavior of the free cantilever where material properties do not affect the dynamics. Here, however, our metrics show that contrast between domains is retrievable by exploring the transients.

Region R1 corresponds to the noncontact approach, where the cantilever displacement is small and the tip–sample interaction is minimal. In R2, the tip snaps into contact with the meniscus, marking the onset of a new attractive force associated with the meniscus formation. Region R3 corresponds to the repulsive contact portion of the interaction where the tip-sample interactions are dominated by mechanical viscoelastic contact between the tip and sample. During R4, the tip withdraws from the surface, and the interaction force decreases. This region is often accompanied by adhesive or meniscus-mediated dynamics. Finally, R5 corresponds to recovery after tip–sample separation, where the cantilever only experiences weak van der Waals interactions and may retain residual oscillatory motion excited by the rapid snap-out transition. In particular, the snap-in and pull-off regions, R2 and R4, contain rapid changes associated with meniscus formation and collapse. R3 contains the contact-dominated response used for polymer stiffness and modulus estimation. The above segmentation provides the framework for the analyses that follow. Each regime can be analyzed separately to extract interaction-dependent observables such as force, stiffness, dissipation, oscillation frequency, and amplitude. In the next sections, we first discuss the broader dynamic behavior observed across R1–R5 and then focus in more detail on two physically important parts of the cycle: the formation and collapse of the meniscus during R2 and R4, and the mechanical contact response during R3.

Conventional force measurements make use of the smoothed displacement signal $\langle z_{\text{tip}} \rangle$, low-pass filtered at a frequency much lower than the natural resonance, typically a few kHz.[2] The quantity $\langle z_{\text{tip}} \rangle$ can be used to fit various nanomechanical models and , by subtracting from the base motion of the cantilever, estimating the indentation of the tip into the surface. Figures 2b-d show the results of using filtered interferometric data to estimate different tip-sample mechanical properties. In addition to this standard analysis, the low noise, high bandwidth signals available from our interferometer data provides additional nanoscale tip-sample interaction information, specifically through tip oscillations that are driven both by stochastic Brownian impulses and by transient impulses originating from rapid changes in the tip-sample boundary conditions.

To explore this, we analyze the oscillatory displacement using single cycle resolved estimators of both the amplitude $A_n$ and frequency $f_n$, where the index n refers to a particular cycle. Briefly, local maxima and minima are detected using prominence-based criteria and are refined by parabolic interpolation. The local oscillation-frequency estimate is obtained from the interval between successive maxima $t_{p,n}$ and minima $t_{v,n}$, whereas the amplitude is calculated from each maximum to the following minimum. These quantities are referred to below as single-cycle amplitude

$$A_n = z_{\text{tip}}(t_{p,n}) - z_{\text{tip}}(t_{v,n}), \qquad (1)$$

and frequency

$$f_n = \frac{1}{t_{p,n+1} - t_{p,n}}. \qquad (2)$$

Since there can be both noise and higher mode signals, we limit the returned oscillation frequency to be in the range $f_{free}/N < f_n < Nf_{free}$, where $f_{free}$ is the free resonance frequency (typically in the range of 60-75 kHz for the levers we used here) and N is a bandwidth factor that excludes low and high frequency contributions, typically chosen to be 2–3. In the following analysis, we returned NaN values for estimates outside the boundary. An example of this is shown in Supplemental Figure S1.

These metrics allow direct comparison between experimental and simulated FC dynamics. Direct observation of these high-bandwidth oscillations addresses a longstanding limitation of ambient force-curve measurements. Conventional displacement detection, detector noise and finite bandwidth can obscure thermally driven and transition-induced cantilever motion.[19] Thermal-noise spectroscopy has long been used to determine resonance frequency, quality factor and spring constant from stationary power spectra.[20] Time-averaged thermally excited contact resonances have similarly been used to infer local elastic, viscoelastic and anelastic properties. [17] Passive overtone microscopy analyzes thermomechanical torsional or antisymmetric modes to estimate lateral or shear contact stiffness. [21] The present work extends this concept to individual fast force curves, in which the interaction state may change within even a single oscillation period. Previous transient-force-curve studies required extensive averaging to recover these dynamics.[9]

**New Information Channels: Connecting Amplitude and Frequency Observables to Physics**

**Brownian amplitude estimates**

Under sufficiently stationary conditions, thermal fluctuations provide a passive stiffness-sensitive signal; near abrupt interaction transitions, excess transient motion reports the release or redistribution of mechanical energy. Changes in resonance frequency provide a convenient measure of the effective normal dynamic stiffness acting on the cantilever. For small frequency shifts, the interaction stiffness can be approximated using a single-mode simple harmonic oscillator (SHO) model, yielding the tip–sample stiffness $k_{ts,SHO}$:

$$k_{ts,SHO} \approx k_c \left[ \left( \frac{f_{\mathrm{res}}}{f_0} \right)^2 - 1 \right]. \qquad (3)$$

While a closed-form relationship enables rapid estimation of stiffness, it becomes inaccurate when the interaction stiffness is large. In that case, a more complete model of the cantilever is required. Here we use an Euler–Bernoulli[22, 23, 24] model to estimate the tip-sample interaction stiffness from the measured resonant frequency. Details are given in the supplemental materials section "Euler-Bernoulli stiffness estimations" (Figures S3–S5). Briefly, a series of tunes were generated by solving the cantilever dynamics over a broad range of trial stiffness values spanning multiple orders of magnitude. For each value of $k_z$, the cantilever response was simulated under harmonic excitation, and the resonance frequency was identified as the first physically meaningful peak in the amplitude–frequency response. This procedure yields a monotonic mapping between stiffness and resonance frequency $k_z \to f_{\text{res}}(k_z)$. This function, in turn, was numerically inverted and interpolated to yield an estimate of the tip-sample stiffness,

$$k_{ts,EB} \approx f_{res}^{-1}(f_{res}). \quad (4)$$

For the data analyzed here, the largest Region 3 frequency estimates at forces below approximately 5 nN are less than 150 kHz. These values correspond to an interaction stiffness of approximately 18 N m−1 in the SHO approximation and 16.5 N m−1 using the Euler–Bernoulli model, a difference of about 10%.

**Brownian- versus transient-dominated cantilever motion**

During a force curve, the cantilever motion contains contributions from two physically distinct sources. Thermal coupling to the environment continuously excites the cantilever, with a characteristic amplitude set by temperature and the effective stiffness of the cantilever–sample system. Abrupt changes in the tip–sample boundary condition—including snap-in, meniscus formation, contact formation, adhesive stretching and pull-off—can additionally inject mechanical energy and generate transient oscillations substantially larger than the equilibrium thermal scale.

To distinguish these regimes, we compare the measured residual oscillation amplitude with the amplitude expected from Brownian motion. Using the contact-resonance stiffness $k_{ts,\text{CR}}$ discussed above, the instantaneous effective stiffness is approximated as $k_z = k_c + k_{ts,\text{CR}}$, where $k_c$is the free cantilever stiffness. From equipartition, the expected mean-square Brownian displacement is $\langle x^2 \rangle = \frac{k_B T}{k_z}$. Thus, the expected Brownian RMS amplitude is $A_{\text{rms,Brown}} = \sqrt{\frac{k_B T}{k_z}}$. Because the measured oscillation amplitude is expressed as a peak-to-peak amplitude, we convert the RMS value to an approximate peak-to-peak Brownian amplitude using

$$A_{pp,\mathrm{Brown}} \approx 2\sqrt{2}\sqrt{\frac{k_B T}{k_z}}. \qquad (5)$$

This conversion treats the local oscillation as approximately sinusoidal over the analysis window. It is therefore not intended as an exact statistical bound on Brownian fluctuations, but rather as a practical normalization for comparing measured oscillation amplitudes across regions with different effective stiffness. We define the Brownian amplitude consistency ratio as

$$R_A = \frac{A_{pp,\mathrm{meas}}}{A_{pp,\mathrm{Brown}}}, \qquad (6)$$

Here, $A_{pp,\mathrm{meas}}$ is the measured peak-to-peak residual amplitude and $A_{pp,\mathrm{Brown}}$ is the peak-to-peak scale obtained from equipartition using the estimated effective stiffness. Values of the Brownian amplitude consistency ratio, RA, of order unity indicate that the observed amplitude is compatible with the thermal scale. Because RA is calculated from short, local peak-to-peak observations rather than from a stationary ensemble average, it is used as a practical consistency indicator rather than a strict pointwise test of equipartition. Persistent values substantially greater than two indicate excess oscillatory motion relative to the thermal expectation and are consistent with transient excitation (see the supplemental materials section “Equipartition Validity Test” and Supplemental Figure S6).

Transient-dominated motion is not simply a larger thermal fluctuation. Rapid boundary-condition changes place the cantilever in a non-stationary state in which the observed motion depends on the recent interaction history. Over these short intervals, time-translation invariance and equilibrium fluctuation–dissipation interpretations need not apply. The measured oscillation therefore reflects both the evolving local stiffness and energy injected or dissipated during the preceding tip–sample transition.

The force-curve regions can therefore be compared using RA, but the classification must be interpreted statistically. In regions where the regime-averaged RA remains of order unity—such as the pre-contact region or portions of stable repulsive contact—the amplitude is broadly compatible with thermal forcing after accounting for the estimated effective stiffness. In regions with persistently elevated RA—such as meniscus formation or post-detachment ringing—the measured motion contains excess energy that cannot be explained by the stiffness-dependent thermal scale alone.

Thus, RA provides a compact normalization for comparing oscillation amplitudes across interaction states with very different stiffnesses. It does not establish thermal equilibrium from a single oscillation; rather, it identifies regimes whose amplitude statistics are compatible with, or substantially exceed, the expected thermal scale. In this framework, the frequency-derived stiffness serves both as a mechanical contrast channel and as the normalization required to identify non-thermal transient excitation.

**Regions R1 and R5: van der Waals regions**

Regions R1 and R5, labeled in Figure 2 correspond to the long-range, non-contact portions of the force curve, where the dominant tip–sample interaction is the relatively weak van der Waals attraction. In R1, before meniscus formation and mechanical contact, the cantilever has already spent roughly half a force curve cycle out of strong contact with the sample and thus might be thermally equilibrated with its environment. This picture was generally consistent with experimental data, as illustrated, for example, in Figure 3**a**. In contrast, R5 occurs after meniscus rupture on retraction. Although the tip has returned to a nominally non-contact state, the cantilever may carry additional kinetic energy released during the snap-off event associated with liquid-bridge rupture. Thus, R5 can exhibit a larger residual oscillatory response than R1, even though both regions are governed primarily by long-range attractive forces (Supplemental Figure S2). The above interpretation is consistent with the experimental results shown in Figures 3a and 3b for the signal $\langle \boldsymbol{A_n} \rangle$ for regions R1 and R5 where the transition between PS/PCL is resolved only in R5. In the latter region, larger $\langle \boldsymbol{A_n} \rangle$ values are obtained for the PCL domain. Consequently, the convergence of $\langle A_n \rangle$ in region R1 points to similar near-field force profiles during approach, while the material transition resolved exclusively in R5 indicates that distinct, asymmetric transient forces govern the retraction phase over the PS and PCL domains.

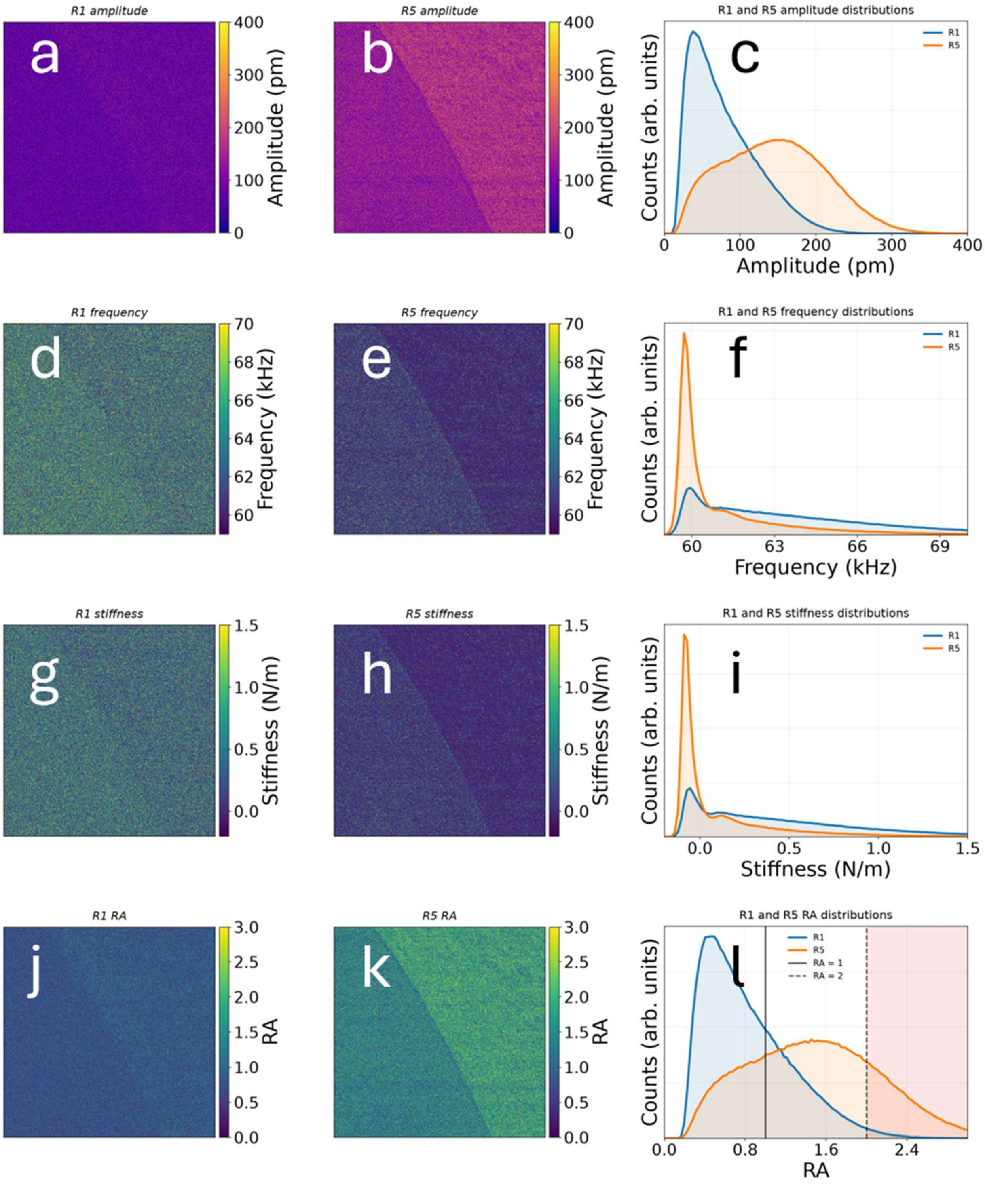

*Figure 3. Region-resolved comparison of transient cantilever dynamics in the non-contact approach region (R1) and the post-detachment recovery region (R5). Maps show the residual oscillation amplitude (a,b), instantaneous frequency (d,e), effective stiffness inferred from the frequency shift (g,h), and Brownian amplitude consistency ratio $R_A$(j,k) for R1 and R5. The corresponding pixel-wise distributions are shown in the right column for amplitude (c), frequency (f), stiffness (i), and $R_A$(l). Matched color scales and histogram axes are used within each row to enable*

*direct comparison between regions. The vertical lines in panel l mark $R_A = 1$, corresponding to the equipartition-based Brownian expectation, and $R_A = 2$, used here as the threshold above which the measured amplitude is larger than expected for Brownian motion alone. The shaded region indicates $R_A > 2$.*

Figure 3 compares the transient cantilever dynamics in the free-oscillation regions before and after the meniscus regions (R2 and R4). Region R1 captures the cantilever motion before the tip encounters the surface and after it has had more time for any meniscus release transition-driven oscillations to relax. As is apparent comparing Figure 3a and b, along with the associated RA maps in Figure 3j and k, the approach amplitude can act a reference for the thermal Brownian motion of the free cantilever. Region R5 occurs after pull-off, once the tip has detached from the surface. Although the cantilever is again nominally free in R5, it has just passed through adhesive and capillary interactions, so its motion can retain more unrelaxed energy injected during detachment.

The amplitude maps in panels a and b show that the residual oscillations are much larger in R5 than in R1. This difference is also clear in the amplitude distributions in panel c, where the R5 histogram is shifted to higher amplitudes and is substantially broader than the R1 distribution. Because the two regions are analyzed with the same estimator and displayed on matched color and histogram scales, the contrast reflects a real change in the cantilever dynamics rather than a visualization artifact.

The instantaneous frequency maps in panels d and e show a corresponding shift toward lower frequencies in R5. Since the local effective stiffness is estimated from the frequency shift, the stiffness maps in panels g and h show the expected reduction in apparent stiffness after detachment;

the corresponding frequency and stiffness distributions in panels f and i show the same systematic shifts. These results indicate that immediately after pull-off, the cantilever is not simply returning to the same free-oscillation state observed before contact. Instead, it oscillates in a dynamically softened state, consistent with the recent release of adhesive and capillary boundary conditions. The implication is that these transients in R5 contain distinctive information that the ergodic mean does not encode, i.e., R5 $\neq$ R1.

The final row tests whether the measured amplitudes are consistent with Brownian motion. The ratio $R_A$compares the measured peak-to-peak residual amplitude with the thermal expectation calculated from equipartition using the local effective stiffness. In R1, the $R_A$map in panel j and histogram in panel l are concentrated near $R_A = 1$, indicating that the pre-contact motion is largely Brownian-consistent. In contrast, R5 shifts to much larger $R_A$values, with many pixels exceeding the $R_A = 2$threshold. These elevated values cannot be explained simply by the lower post-detachment stiffness, because that stiffness is already included in the thermal amplitude estimate. The excess $R_A$ therefore indicates that additional mechanical energy is transferred to the cantilever during pull-off.

**Regions R2 and R4: Capillary regions**

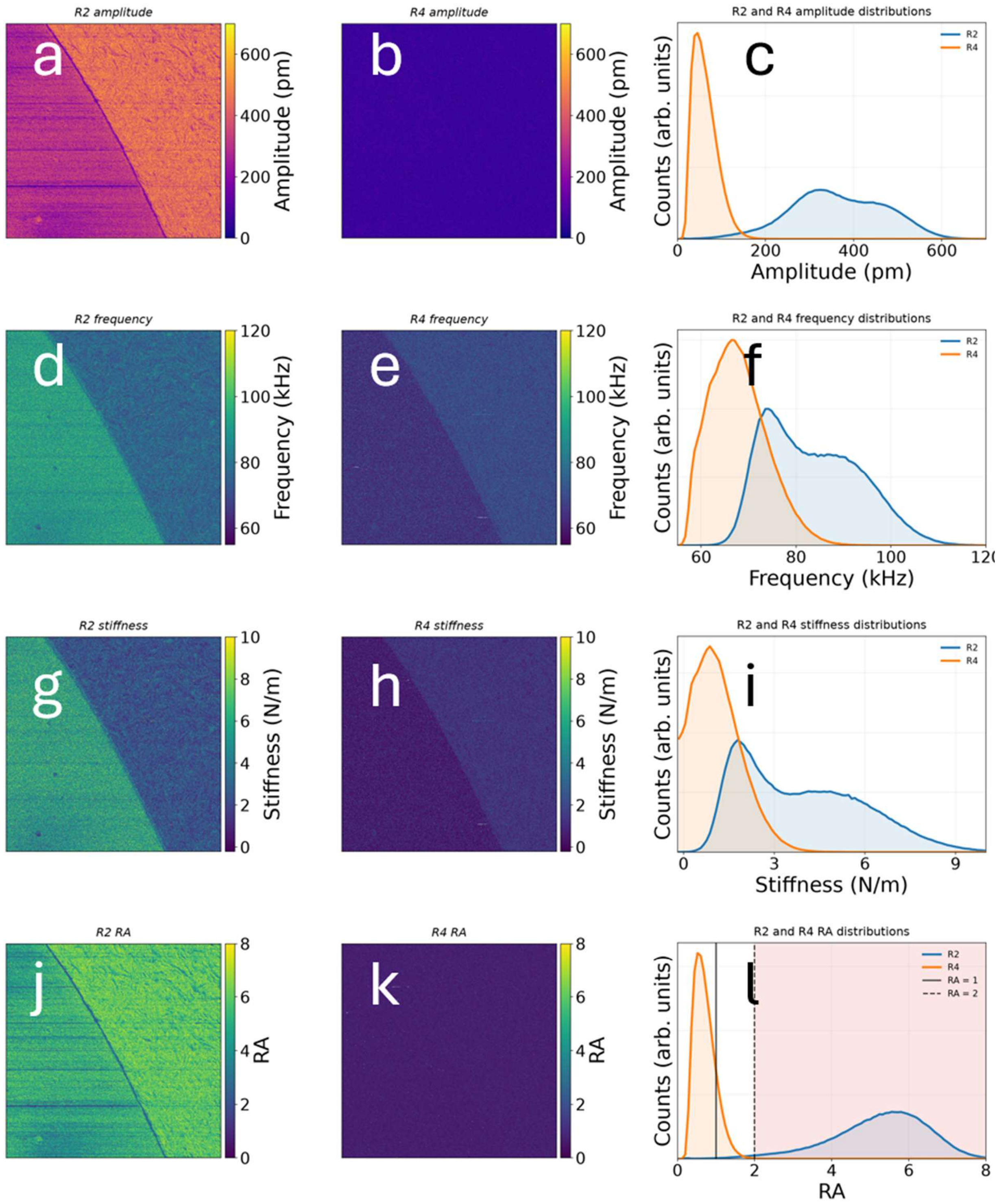

*Figure 4. Region-resolved comparison of transient cantilever dynamics during meniscus-mediated approach and withdrawal. Maps show the residual oscillation amplitude (a,b), instantaneous frequency (d,e), effective stiffness inferred from the frequency shift (g,h), and Brownian amplitude consistency ratio $R_A$(j,k) for the approach meniscus region R2 and the withdrawal meniscus region R4. The corresponding pixel-wise distributions are shown in the right*

*column for amplitude (c), frequency (f), stiffness (i), and $R_A$(l). Matched color scales and histogram axes are used within each row to enable direct comparison between the two regions. The vertical lines in panel l mark $R_A = 1$, corresponding to the equipartition-based Brownian expectation, and $R_A = 2$, used here as the threshold above which the measured amplitude is larger than expected for Brownian motion alone. The shaded region indicates $R_A > 2$, where the oscillation amplitude exceeds the thermal prediction.*

Unlike the nominally free regions R1 and R5, regions R2 and R4 both occur while the cantilever is coupled to the sample through a liquid meniscus. However, these two meniscus-mediated states are not dynamically equivalent. R2 corresponds to meniscus formation during approach, when the liquid bridge first establishes a mechanical connection between the tip and surface. R4 corresponds to withdrawal, when the bridge is stretched and progressively weakened before rupture. The comparison in Figure 4 shows that this hysteresis produces a strong asymmetry in the cantilever response.

The amplitude maps in panels a and b show substantially larger residual oscillation amplitudes in R2 than in R4. This contrast is also evident in the amplitude distributions in panel c: the R2 distribution is shifted to much higher amplitudes and is much broader, whereas R4 is concentrated at low amplitudes. The larger R2 amplitudes indicate that meniscus formation acts as a strong transient excitation event. By contrast, during R4 the cantilever remains coupled through the stretched meniscus, but the measured oscillations are much smaller, suggesting that the withdrawal branch is mechanically quieter until final rupture.

The instantaneous frequency maps in panels d and e show that R2 also has higher oscillation frequencies than R4. Since the frequency shift is converted into an effective stiffness, the maps in panels g and h and the distributions in panels f and i show the same trend: R2 exhibits a substantially larger effective stiffness, while R4 is shifted toward lower stiffness. This behavior is consistent with a newly formed meniscus acting as a relatively strong restoring constraint during approach. During withdrawal, the liquid bridge is extended and closer to instability, so it provides a weaker effective constraint before detachment.

The Brownian amplitude consistency ratio $R_A$ provides an additional test of whether the measured oscillations can be explained by thermal motion alone. In R4, the $R_A$ map in panel k and distribution in panel l are concentrated near unity and lie mostly below the $R_A = 2$ threshold, indicating that the withdrawal-region oscillations are largely consistent with Brownian motion once the local stiffness is accounted for. In contrast, R2 (panel j) is shifted to much larger $R_A$ values, with most pixels exceeding the Brownian-consistent range. Because the equipartition-based thermal amplitude estimate already includes the local effective stiffness, the elevated R2 $R_A$ values cannot be explained simply by the higher stiffness in R2. Instead, they show that additional mechanical energy is injected into the cantilever during meniscus formation.

These measurements demonstrate that the approach and withdrawal portions of the capillary interaction are strongly asymmetric. Although both regions involve a liquid bridge, R2 is characterized by larger oscillation amplitudes, higher instantaneous frequencies, larger effective stiffness, and a much stronger departure from Brownian equilibrium than R4. This indicates that meniscus formation excites the cantilever more strongly than meniscus stretching during withdrawal. The result highlights the hysteretic nature of the capillary interaction and shows that transient dynamic analysis provides contrast that is not available from quasi-static force measurements alone.

**Region R3: Contact Mechanical region**

In the following section, we compare two estimates of the local stiffness in Region 3, the elastic-contact portion of the force curve: (1) a conventional force–indentation analysis and (2) a contact-resonance analysis. Standard force-curve analysis and contact-mechanics assumptions are reviewed elsewhere.[2]

### a. Conventional force curve analysis

A typical analysis of a force curve stiffness can be made by evaluating the applied force versus indentation into the sample surface. Details of this standard analysis are described in the supplemental materials section "Sub-resonant, displacement based stiffness estimates". Briefly, three contact-mechanics models were then applied to the filtered force-indentation data: punch, Hertz, and DMT. Under the punch approximation, force was modeled as linearly proportional to indentation, $F = S_{\mathrm{p}}\delta$, and the mapped punch stiffness was taken directly as the fitted slope $S_{\mathrm{p}}$. For the Hertz model, force was fit as $F = S_{\mathrm{H}}\delta^{3/2}$, corresponding to spherical contact with tip radius $R$. The fitted coefficient was converted to an effective modulus through $E_{\mathrm{H}} = 3S_{\mathrm{H}}/(4\sqrt{R})$, and a local stiffness evaluated at the median positive indentation in Region 3, $\delta_{\mathrm{med}}$, using $K_{\mathrm{H}} = 2E_{\mathrm{H}}\sqrt{R\delta_{\mathrm{med}}}$ (Figure 6a). The DMT model was implemented as an adhesion-corrected Hertz form, $F + F_{\mathrm{adh}} = S_{\mathrm{DMT}}\delta^{3/2}$, with the corresponding modulus $E_{\mathrm{DMT}} = 3S_{\mathrm{DMT}}/(4\sqrt{R})$and local stiffness $K_{\mathrm{DMT}} = 2E_{\mathrm{DMT}}\sqrt{R\delta_{\mathrm{med}}}$ (Figure 6b). In the adhesion-based mode used here, the adhesion force was estimated directly from the retraction segment as the magnitude of the minimum baseline-corrected force.

The adhesion force entering the DMT model was taken from the most negative force during the retraction regime. With the cantilever treated as a spring, this corresponds to the minimum value of $F(t) = k_c w(t)$ in Region 4. This produces both the DMT correction term and the separate adhesion map shown in Figure 2c.

### b. Contact resonance and tip-sample stiffness.

The contact-resonance frequency of a tip pressed into a surface is dependent on the normal tip-sample stiffness $k_{ts}$. This stiffness was estimated from the frequency shift as discussed above (see Equations (3) and (4)).

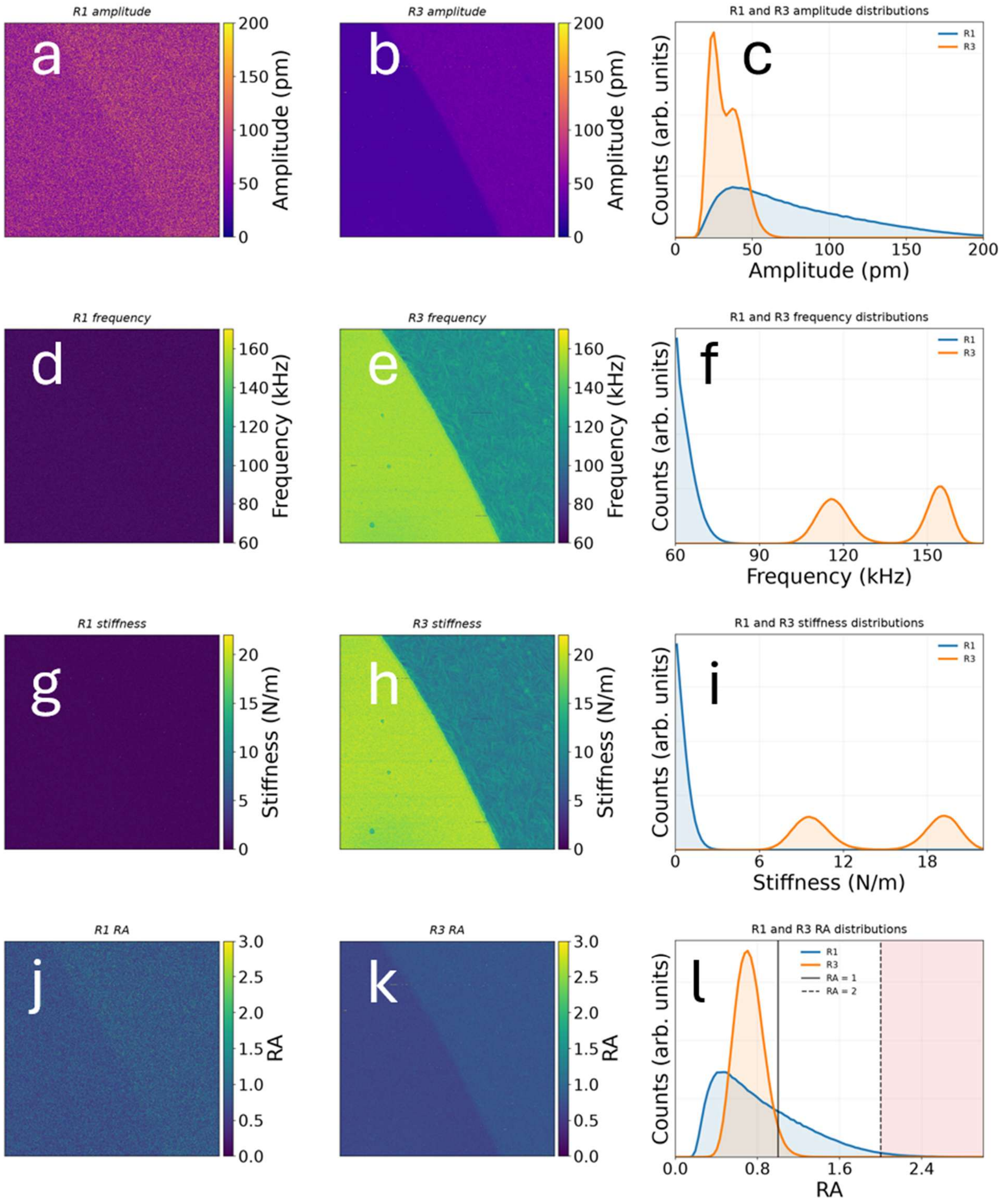

***Figure 5. Region-resolved comparison of transient cantilever dynamics before tip–sample interaction and during repulsive contact.*** *Maps show the residual oscillation amplitude (a,b), instantaneous frequency (d,e), effective stiffness inferred from the frequency shift (g,h), and Brownian amplitude consistency ratio* $R_A$*(j,k) for the pre-contact non-contact region R1 and the repulsive contact region R3. The corresponding pixel-wise distributions are shown in the right column for amplitude (c), frequency (f), stiffness (i), and* $R_A$*(l). Matched color scales and histogram axes are used within each row to enable direct comparison between the free cantilever state and the contact-stiffened state. The*

*frequency and stiffness maps show the large upward resonance shift and increased effective stiffness that occur when the tip is mechanically coupled to the sample in R3. The vertical lines in panel l mark* $R_A = 1$*, corresponding to the equipartition-based Brownian expectation, and* $R_A = 2$*, used here as the threshold above which the measured amplitude is larger than expected for Brownian motion alone. The shaded region indicates* $R_A > 2$*. The R3* $R_A$*distribution remains largely below this threshold, indicating that the reduced contact-region amplitude is broadly consistent with Brownian motion after accounting for the increased local stiffness.*

Figure 5 shows that the transient oscillations observed during repulsive contact provide quantitative dynamic information that complements conventional force-curve analysis. The figure compares the pre-contact free-oscillation region R1 with the repulsive contact region R3, where the tip indents the sample and the cantilever is mechanically coupled to the surface. Unlike the capillary-dominated regimes R2 and R4, R3 corresponds to direct tip–sample contact, so the cantilever dynamics are governed primarily by the local contact stiffness.

The amplitude maps in panels a and b show that the residual oscillation amplitude is reduced during contact. This reduction is also evident in the distributions in panel c, where the R3 histogram is shifted to lower amplitudes and is narrower than the R1 distribution. This behavior is expected because contact increases the effective stiffness of the cantilever–sample system. A stiffer boundary condition suppresses displacement fluctuations, so the thermally driven oscillation amplitude decreases during repulsive indentation.

The instantaneous frequency maps in panels d and e show the complementary effect. In R3, the oscillation frequency increases strongly relative to the free-cantilever response in R1. This upward frequency shift reflects the additional stiffness introduced by the repulsive tip–sample contact. When the measured frequency is converted into effective stiffness, the resulting maps in panels g and h show clear spatial contrast between the two polymer domains, with the corresponding distributions shown in panels f and i. The close correspondence between the R3 frequency and stiffness maps confirms that the transient resonance analysis is sensitive to local contact stiffness during the indentation portion of each force curve.

The Brownian amplitude consistency ratio $R_A$ provides an independent check on whether the measured amplitudes are consistent with thermal fluctuations after accounting for the contact-induced stiffness increase. Although the absolute oscillation amplitude is much smaller in R3, the $R_A$ maps in panels j and k and distribution in panel l remain largely below the $R_A = 2$ threshold. This indicates that the reduction in amplitude is mostly explained by the increased contact stiffness through the equipartition-based thermal amplitude estimate. In other words, R3 is not characterized by the large excess excitation seen during meniscus formation in R2. Instead, the repulsive contact region behaves more like a thermally driven oscillator constrained by a stiffer mechanical boundary condition.

Thus, R3 is dynamically distinct from R1 as the large contact stiffness shifts the resonance upward and reduces the Brownian amplitude. The combined amplitude, frequency, stiffness, and $R_A$ measurements show that transient oscillations during contact can provide quantitative contact-stiffness contrast while also testing whether the measured motion remains consistent with thermal equilibrium under the local contact boundary condition.

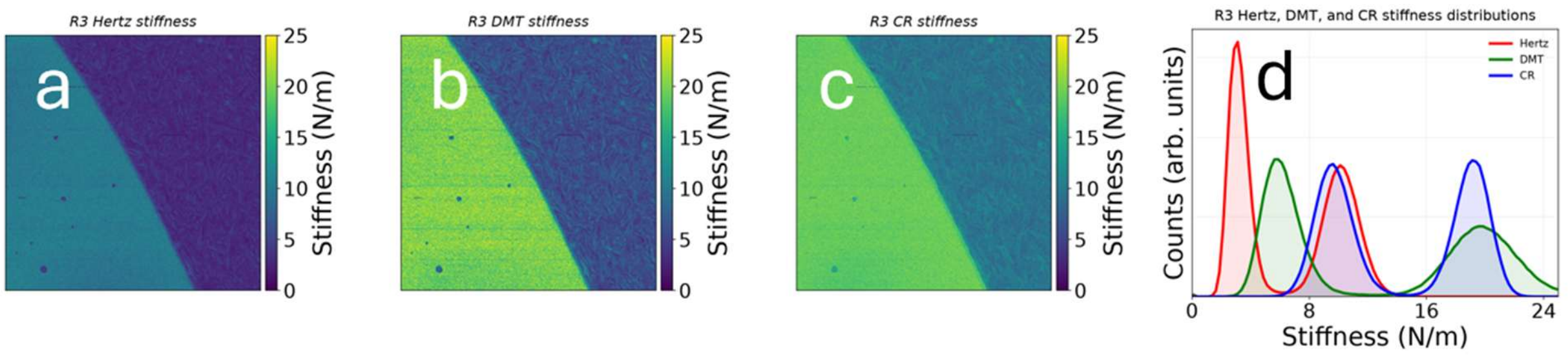

*Figure 6. Comparison of contact-region stiffness estimates obtained from contact mechanics and transient resonance analysis. Maps show the R3 stiffness calculated from Hertz contact mechanics (a), DMT contact mechanics (b), and the transient contact-resonance response (c). All three maps are displayed with the same color scale to enable direct comparison of spatial contrast and absolute stiffness values. The corresponding pixel-wise stiffness distributions are shown in panel d. The Hertz, DMT, and CR maps all recover the same two-domain material contrast, but the distributions differ in absolute stiffness and peak separation, reflecting the different assumptions and physical sensitivities of the three approaches. Hertz and DMT stiffness are derived from quasi-static contact-mechanics models, whereas CR stiffness is inferred dynamically from the contact-induced resonance-frequency shift during R3.*

Figure 6 compares three estimates of the R3 contact stiffness obtained from contact mechanics and transient contact-resonance analysis. The Hertz and DMT stiffness maps in panels a and b are derived from quasi-static force–indentation analysis, whereas the CR stiffness map in panel c is inferred directly from the transient resonance frequency during repulsive contact. All three methods recover the same broad spatial contrast between the PS and PCL domains, confirming that the R3 frequency shift is strongly coupled to local contact stiffness. However, the absolute stiffness values differ substantially between the models.

The Hertz stiffness estimate disagrees most strongly with the CR stiffness. In the histogram comparison in panel d, the Hertz distribution is shifted to much lower stiffness than the CR distribution and does not reproduce the CR peak positions, particularly for the stiffer PS domain. This indicates that a purely elastic, non-adhesive Hertz model does not adequately describe the contact mechanics in this measurement. By neglecting adhesion, the Hertz model underestimates the effective mechanical constraint experienced by the cantilever during contact.

In contrast, the DMT stiffness agrees much more closely with the CR stiffness, especially for the PS domain. The DMT and CR maps show similar spatial structure, and their high-stiffness histogram peaks are much better aligned than the corresponding Hertz peak. This improved agreement is significant because the DMT model includes adhesive interactions while retaining an elastic contact framework. The comparison therefore suggests that adhesion makes an important contribution to the effective contact stiffness during R3.

In this sense, the CR stiffness measurement has effectively "discovered" the importance of adhesion without assuming a particular contact-mechanics model in advance. The CR approach measures the dynamic stiffness through the cantilever's resonant response, while the Hertz and DMT estimates impose different quasi-static contact assumptions. The fact that the CR result aligns more closely with DMT than Hertz indicates that the contact resonance is sensing an adhesive contact boundary condition, not a purely Hertzian elastic indentation. Thus, the transient resonance analysis provides not only a dynamic stiffness map, but also a way to test which contact-mechanics assumptions are most consistent with the measured cantilever dynamics.

## Discussion

High-bandwidth QPDI measurements show that individual fast force curves contain both ever-present thermal fluctuations and event-driven transient oscillations. Segmenting the modulation cycle into five interaction regimes reveals systematic differences in amplitude and cycle-resolved frequency that are spatially correlated with the PS and PCL domains. A state-dependent single-mode model reproduces the principal qualitative trends and provides a framework for relating them to changes in stiffness, capillary coupling and boundary-condition transitions.

The strongest non-stationary responses occur near snap-in, meniscus formation and pull-off. These events abruptly change the equilibrium position and mechanical boundary condition of the cantilever and can inject energy into its flexural motion. In contrast, the comparatively stable portions of R1 and R3 exhibit amplitudes that are broadly compatible with the stiffness-dependent thermal scale. The distinction is not absolute at the level of a single oscillation: the Brownian amplitude consistency ratio is a statistical diagnostic, and elevated values identify excess motion without uniquely determining the microscopic excitation or dissipation mechanism.

A peak-and-trough estimator was used because the relevant oscillations often persist for only a few cycles and change frequency within the analysis interval. Windowed spectral and Hilbert-transform methods are less reliable under these strongly non-stationary conditions, whereas cycle-resolved estimates retain the local timing of the response. This advantage is accompanied by greater statistical scatter and sensitivity to peak selection. Frequency and amplitude should therefore be interpreted together with the raw trace, estimator bounds and spatial distributions rather than as exact instantaneous state variables.

The per-pixel analysis preserves heterogeneity that would be suppressed by registration and averaging. Spatial variations in frequency, amplitude and decay can reflect differences in effective stiffness and in the energy transferred during capillary or adhesive transitions. Amplitude decay alone is not a unique measure of dissipation, because it is also affected by changes in stiffness, equilibrium position, mode coupling and detector bandwidth. Quantitative damping or loss measurements will require explicit fitting of the transient response or an independently validated dynamical model.

We emphasize that averaged curves, for example, those reported by Eppell et al. are qualitatively different from the single, time-resolved QPDI force curves reported here, despite being measured on similar materials with similar operating parameters. Formally, ergodicity implies that the time average of an observable *o(t)* over a single, sufficiently long trajectory equals the ensemble average over all accessible microstates Γ, weighted by their equilibrium probability ρ(Γ):

$$\lim_{T\to\infty} \frac{1}{T}\int_0^T o(t)\, dt \equiv \langle o \rangle_{ensemble} = \int o(\Gamma)\, \rho(\Gamma)\, d\Gamma \qquad (7)$$

On the other hand, non-ergodicity for a given measurement, mechanism or phenomena implies that

$$\frac{1}{T}\int_0^T o(t)\, dt \neq \langle o \rangle_{ensemble} \quad (8)$$

When the inequality in Eq. (8) persists as T→∞, it describes strictly non-ergodic phenomena—mechanisms comprising memory-dependent states, such as the approach and retraction branches of a capillary bridge. In such cases, an average over the branches can still be constructed, but it is artificial: the true material properties are encoded in each branch separately. A mean displacement retrieved for each individual branch contains state-dependent information that is absent from the artificial average.

We now formalize the method, introduced above, that deliberately breaks ergodicity, which we term *measurement-broken ergodicity*: measuring with a time window shorter than that required to recover the mean, when such a mean exists. As a trivial example, ergodicity is broken by measuring the response of a simple harmonic oscillator before a transient force has relaxed; if the transient response is large enough, the ergodic mean cannot be recovered. Notably, even for genuinely ergodic mechanisms such as Brownian fluctuations, the inequality in Eq. (8) holds for any finite observation time T. Averaging then preserves the reproducible components of the response, per Eq. (7), because the underlying system remains ergodic—but it suppresses stochastic thermal motion and cycle-to-cycle variability.

There are several scenarios in which material properties can be recovered *only* by breaking ergodicity through measurement, and here we extend this concept to the individual branches of non-ergodic mechanisms such as the capillary bridge. The first scenario involves short-lived events whose decay contains material-specific information. Although a mean remains meaningful and recoverable, it is silent on the kernel shape of the decaying event: all of that information resides in the transients.[25] The second scenario involves events that repeat only a finite number of times. Here there is no ensemble to average over, since the information is encoded in that event, or set of events, itself.

For these reasons, even though a mean can be recovered for mechanisms involving capillary formation and rupture, atomic hopping, stick-slip, stretching or unfolding of polymer chains, charge trapping, ferroelectric switching, or domain depinning—and while such a mean is sometimes useful—it may (1) yield material properties quantified from an artificial average, (2) miss a material property entirely by missing a tip trajectory, or (3) discard non-repeatable transient information that the mean does not encode.

The present approach is complementary to intermittent-contact resonance AFM, driven contact-resonance methods and ringing-mode analysis. Intermittent-contact resonance AFM synchronizes high-speed resonance tracking with the applied force during each intermittent contact.[26, 27] Driven methods provide high signal-to-noise ratio and controlled excitation, but their interpretation can depend on the actuator transfer function.[28, 29] Photothermal actuation reduces some of these complications, although frequency-dependent transfer-function effects remain relevant.[30, 31, 32] Thermal excitation avoids an applied resonant drive and can probe the contact with very small oscillation amplitudes, providing a direct connection to fluctuation-based mechanical response.[33, 34] Transition-induced ringing, by contrast, provides sensitivity to abrupt changes in adhesion and capillary state and is closely related to ringing-mode analysis.[15] The cost is reduced statistical precision and a stronger requirement for low detector noise, high acquisition bandwidth and careful separation of stationary fluctuations from non-stationary transients.

Several limitations define the present quantitative scope. The analysis assumes that a locally dominant flexural mode can be represented by a cycle-resolved frequency and converted to an effective normal dynamic stiffness using a prescribed cantilever model. The short signals limit the precision with which frequency and damping can be separated. The Brownian-consistency ratio is a practical statistical indicator rather than a direct pointwise test of equipartition. Conversion from stiffness to modulus additionally depends on tip radius, contact geometry, adhesion model and possible rate dependence of the polymer. Broader validation will require systematic measurements as a function of humidity, force-curve rate, loading force, cantilever stiffness and sample viscoelasticity.

The post-detachment dynamics are closely related to ringing-mode measurements, but the present force curves use comparatively small modulation ranges of approximately 10–40 nm. At fixed modulation frequency, reducing the range shortens the time spent in the interaction and can limit the development of rate-dependent adhesive processes such as capillary-bridge growth, polymer-chain extension, viscoelastic relaxation and hydration-layer restructuring. The stored adhesive energy and the impulsiveness of detachment may therefore decrease, producing weaker pull-off signatures and smaller post-snap-off oscillations.

From an energy-partition perspective, the ringing amplitude depends on the fraction of interaction energy transferred into cantilever motion. A smaller or more gradual detachment can dissipate energy locally at the interface rather than converting it into flexural kinetic energy. Conversely, longer-range or slower force curves may allow a capillary or adhesive junction to develop more fully and can increase the energy released at rupture. The ringing amplitude should therefore be interpreted as a rate- and history-dependent observable rather than as a direct measure of adhesion alone.

Future experiments that vary modulation range and rate independently should test this interpretation by comparing pull-off work, rupture distance, post-detachment amplitude and decay. Such measurements would help separate the roles of capillary kinetics, polymer viscoelasticity and cantilever dynamics.

## Conclusion

High-bandwidth interferometric detection reveals that fast AFM force curves contain a dynamic response that is largely removed by conventional filtering and averaging—a signal that has remained essentially hidden in the forty years since the invention of the AFM, apart from the large oscillations associated with relatively large adhesive separations (snap from contact) and time-averaged approaches with their own inherent challenges. Within an individual modulation cycle, the cantilever exhibits both ever-present thermal fluctuations and short-lived oscillations generated by abrupt changes in the tip–sample boundary condition. Segmenting each force curve into non-contact approach, meniscus formation, repulsive contact, meniscus-mediated withdrawal and post-detachment recovery makes these contributions spatially and mechanically distinguishable.

The cycle-resolved oscillation frequency provides a stiffness-sensitive observable – the contact resonance frequency - during repulsive contact. On the PS/PCL blend measured here, frequency-derived effective-stiffness maps recover the same material-domain contrast as conventional force–indentation analysis without requiring an added resonant excitation or a separate imaging pass. In contrast, the enlarged oscillation amplitudes observed during meniscus formation and after pull-off exceed the stiffness-dependent thermal scale, consistent with transient mechanical energy being transferred to the cantilever during rapid changes in the capillary and adhesive boundary conditions.

Comparison of experiment and simulation demonstrates both the value and the limitations of reducing the interaction to a single elastic parameter. Parameter sets that reproduce the low-frequency force–indentation response do not necessarily reproduce the measured contact-resonance frequency, whereas parameter sets chosen to match the dynamic frequency can degrade agreement with the quasi-static force curve. The two observables therefore provide complementary constraints on the tip–sample interaction, and their divergence motivates models incorporating adhesion, nonlinear contact geometry, viscoelasticity and state-dependent capillary interactions.

More generally, these results suggest force curves are best understood as a multifrequency measurement rather than solely a quasi-static force–indentation record—and that a substantial amount of accessible information has, until now, either been routinely ignored or discarded. The slowly varying displacement reports the mean interaction force and indentation, while the fluctuations and transients report the evolving mechanical boundary condition on microsecond timescales, all within a single, passive measurement requiring no added excitation or separate imaging pass. With appropriate statistical treatment and calibration, this opens several promising directions: rapid, pixel-resolved mapping of effective stiffness; localization and counting of non-equilibrium events such as meniscus formation, adhesive rupture and contact instabilities; and, perhaps most intriguingly, a new class of non-repeating measurements in which the information resides in a single transient—the nucleation and rupture of an individual capillary bridge, a single stick–slip or atomic-hopping event, the unfolding of an individual polymer chain, an isolated charge-trapping or ferroelectric-switching event, or the depinning of a single domain boundary. Because the cycle-resolved estimators capture the frequency, amplitude and decay of each event without averaging, such one-time phenomena become quantifiable at the pixel level for the first time. More broadly, this hitherto hidden signal may pave the way for new ways of studying and exploiting Brownian motion in a variety of systems: we now have the capability of studying time-resolved (non-)Markovian and (non-)ergodic dynamics mediated by the tip–sample interaction, including single-event dynamics. Realizing this potential will require quantifying estimator uncertainty, testing the frequency dependence of the inferred response, and validating the method systematically across environmental conditions, loading rates and material classes; these are tractable next steps, and the breadth of mechanisms already visible in individual force curves suggests the effort will be well rewarded.

## Methods

Fast force-mapping measurements (FFM) were performed using an AFM (Vero, Asylum Research–Oxford Instruments, Santa Barbara, CA) equipped with high-bandwidth quadrature phase, differential interferometric (QPDI) displacement detection. At each pixel, the instrument recorded synchronized time traces of cantilever displacement and base (z-sensor) motion, stored in HDF5 format. FCs are generated by sinusoidal modulation of the tip–sample separation, typically from 100-2000Hz. The displacement and base motion were measured at 2MHz using the standard FFM mode available on the Vero AFM. The lowpass filter bandwidth was set to 500kHz, allowing unfiltered acquisition of the displacement signal well beyond the first pinned contact resonance frequency of the cantilever.

## Data availability

The data that support the findings of this study are available from the corresponding author upon reasonable request.

## Code availability

Analysis code used to generate the results in this study is available from the corresponding author upon reasonable request.

## References

1 G. Binnig, C. F. Quate, and C. Gerber, Physical Review Letters **56** (9), 930 (1986).

2 H. J. Butt, B. Cappella, and M. Kappl, Surface Science Reports **59** (1-6), 1 (2005).

3 H. J. Butt and M. Kappl, Advances in Colloid and Interface Science **146** (1-2), 48 (2009).

4 B. Rajabifar, Y. M. Jadhav, D. Kiracofe, G. F. Meyers, and A. Raman, MACROMOLECULES **51** (23), 9649 (2018).

5 T. Hugel and M. Seitz, MACROMOLECULAR RAPID COMMUNICATIONS **22** (13), 989 (2001).

6 F. L. Leite, C. C. Bueno, A. L. Da Róz, E. C. Ziemath, and O. N. Oliveira, INTERNATIONAL JOURNAL OF MOLECULAR SCIENCES **13** (10), 12773 (2012).

7 A. RosaZeiser, E. Weilandt, S. Hild, and O. Marti, Measurement Science and Technology **8** (11), 1333 (1997).

8 I. Sokolov and M. E. Dokukin, in *Nanoscale Imaging: Methods and Protocols*, edited by Y. L. Lyubchenko (2018), Vol. 1814, pp. 469.

9 S. J. Eppell, L. Li, and F. R. Zypman, Aip Advances **7** (10) (2017).

10 R. Garcia, Chemical Society Reviews **49** (16), 5850 (2020).

11 R. Garcia and J. R. Tejedor, NANOSCALE ADVANCES **7** (20), 6286 (2025).

12 C. Braunsmann, J. Seifert, J. Rheinlaender, and T. E. Schäffer, REVIEW OF SCIENTIFIC INSTRUMENTS **85** (7) (2014).

13 S. Kim, Y. Lee, M. Lee, S. An, and S. J. Cho, NANOMATERIALS **11** (6) (2021).

[14] M. Li, N. Xi, and L. Q. Liu, NANOSCALE **13** (18), 8358 (2021).

[15] M. E. Dokukin and I. Sokolov, Scientific Reports **7** (2017).

[16] S. Prasad, A. Rankine, T. Prasad, P. Song, M. E. Dokukin, N. Makarova, V. Backman, and I. Sokolov, Advanced Nanobiomed Research **1** (8) (2021).

[17] C. F. Ma, C. G. Zhou, J. L. Peng, Y. H. Chen, W. Arnold, and J. R. Chu, JOURNAL OF APPLIED PHYSICS **129** (23) (2021).

[18] D. C. Lin, E. K. Dimitriadis, and F. Horkay, JOURNAL OF BIOMECHANICAL ENGINEERING-TRANSACTIONS OF THE ASME **129** (3), 430 (2007).

[19] J. Lübbe, M. Temmen, P. Rahe, and M. Reichling, BEILSTEIN JOURNAL OF NANOTECHNOLOGY **7**, 1885 (2016).

[20] A. Gannepalli, A. Sebastian, J. P. Cleveland, M. V. Salapaka, and acc, in *PROCEEDINGS OF THE 2004 AMERICAN CONTROL CONFERENCE, VOLS 1-6* (2004), pp. 3122.

[21] T. Drobek, R. W. Stark, and W. M. Heckl, PHYSICAL REVIEW B **64** (4) (2001).

[22] S. Bradler, S. R. Kachel, A. Schirmeisen, and B. Roling, JOURNAL OF APPLIED PHYSICS **120** (16) (2016).

[23] S. Bradler, A. Schirmeisen, and B. Roling, JOURNAL OF APPLIED PHYSICS **122** (6) (2017).

[24] S. Bradler, A. Schirmeisen, and B. Roling, JOURNAL OF APPLIED PHYSICS **123** (3) (2018).

[25] Jing-Dong Bao, Peter Hänggi, and Yi-Zhong Zhuo, Physical review. E, Statistical, nonlinear, and soft matter physics **72** (6), 061107 (2005).

[26] G. Stan and R. S. Gates, Nanotechnology **25** (24) (2014).

[27] G. Stan and S. W. King, Journal of Vacuum Science & Technology B **38** (6) (2020).

[28] R. Proksch and S. V. Kalinin, Nanotechnology **21** (45) (2010).

[29] A. Labuda, K. Kobayashi, D. Kiracofe, K. Suzuki, P. H. Grütter, and H. Yamada, Aip Advances **1** (2) (2011).

[30] M. Kocun, A. Labuda, A. Gannepalli, and R. Proksch, Review of Scientific Instruments **86** (8) (2015).

[31] R. Wagner and J. P. Killgore, Applied Physics Letters **107** (20) (2015).

[32] A. R. Piacenti, C. Adam, N. Hawkins, R. Wagner, J. Seifert, Y. Taniguchi, R. Proksch, and S. Contera, Macromolecules **57** (3), 1118 (2024).

[33] M. R. Paul, M. T. Clark, and M. C. Cross, Nanotechnology **17** (17), 4502 (2006).

[34] A. Roters, M. Gelbert, M. Schimmel, J. Ruhe, and D. Johannsmann, Physical Review E **56** (3), 3256 (1997).

## Acknowledgements

## Author contributions

S.S. and R.P. conceived the study, analyzed the data, interpreted the results and wrote the manuscript.

## Competing interests

R.P. is an employee of Asylum Research, Oxford Instruments, the manufacturer of the AFM used in this study. S.S. declares no competing interests.